\documentclass[
    aps, prb,
    superscriptaddress, 
    showkeywords,
    reprint,
    nolinenumbers,
    amsmath,amssymb,floatfix.xcolor=table
    ]{revtex4-2}

\usepackage[utf8]{inputenc}
\usepackage{ulem}
\usepackage{amssymb}
\renewcommand*{\vec}[1]{\mathbf{#1}}

\usepackage[T1]{fontenc} % show | as vertical line...
\usepackage{graphicx}
\usepackage[separate-uncertainty=true]{siunitx}

\usepackage{xcolor}

\DeclareUnicodeCharacter{2215}{\\} 

\usepackage{hyperref}

\makeatletter
\def\maketitle{
	\@author@finish
	\title@column\titleblock@produce
	\suppressfloats[t]}
\makeatother

\newcommand{\beginsupplement}{% redefine counters for supplement
		\setcounter{page}{1}
		\setcounter{section}{0}
		\renewcommand{\thesection}{S\arabic{section}}%
		\setcounter{table}{0}
		\renewcommand{\thetable}{S\arabic{table}}%
		\setcounter{figure}{0}
		\renewcommand{\thefigure}{S\arabic{figure}}%
		\setcounter{equation}{0}
		\renewcommand{\theequation}{S\arabic{equation}}%
		}

\newcommand{\unitokyo}{Institute for Solid State Physics, University of Tokyo, Kashiwa 277-8581, Japan}
\newcommand{\iitm}{Department of Physics, Indian Institute of Technology Madras, Chennai 600036, India}
\newcommand{\riken}{Center for Emergent Matter Science, RIKEN, 2-1 Hirosawa, Wako 351-0198, Japan}
\newcommand{\nanogune}{CIC nanoGUNE BRTA, 20018 Donostia-San Sebasti\'{a}n, Spain}
\newcommand{\lboro}{Department of Physics, Loughborough University, Epinal Way, Loughborough, LE11 3TU, UK}
\newcommand{\epfl}{École polytechnique fédérale de Lausanne (EPFL), School of Engineering, Institute of Materials, Laboratory of Nanoscale Magnetic Materials and Magnonics, 1015 Lausanne, Switzerland}
\newcommand{\fraunhofer}{Fraunhofer ITWM, 67663 Kaiserslautern, Germany}

\newcommand{\papertitle}{Electrical detection of interfacial exchange field at the (ferromagnetic insulator)$|$(normal metal) interface using spin-dependent scattering}

\begin{document}
	
	%\newrefsection % separate bibliography for main text and supplement
	
	\title{\papertitle}
	
	\author{Prasanta Muduli $^{\ddag}$}
	\thanks{Contributed equally}
	\email{muduli@iitm.ac.in}
	% ORCID 0000-0001-8568-9879
	\affiliation{\unitokyo}
	\affiliation{\iitm}

	\author{Na\"{e}mi Leo $^{\ddag}$}
	\thanks{Contributed equally}
	\email{n.leo@lboro.ac.uk}
	% ORCID 0000-0002-0828-6889
	\affiliation{\unitokyo}
	\affiliation{\nanogune}
	\affiliation{\lboro}
	
	%\thanks{when we publish I probaby will be at TUWien...}
	
	\author{Mingran Xu}
	% ORCID is “0000-0002-0397-6629”
	\affiliation{\unitokyo}
	\affiliation{\riken}
	\affiliation{\epfl}
	
	\author{Zheng Zhu}
	% ORCID ???
	\affiliation{\unitokyo}
	
	\author{Jorge Puebla}
	% ORCID "0000-0002-4364-5672"
	\affiliation{\riken}
	
	\author{Christian Ortiz Pauyac}
	% ORCID?
	\affiliation{\fraunhofer}
	
	\author{Hironari Isshiki}
	% ORCID "0000-0001-8421-4484"
	\affiliation{\unitokyo}
	
	\author{YoshiChika Otani} 
	% ORCID "0000-0001-8008-1493"
	\affiliation{\unitokyo}
	\affiliation{\riken}

	\begin{abstract}
		
The spin-orbit field and interfacial exchange field are two major interface phenomena, and the detection and manipulation of these fields can enable a variety of nanoscale spintronics devices. 
Optimizing the interfacial exchange field, which governs the spin-dependent scattering asymmetry at (ferromagnetic insulator)-(normal metal) interfaces, will pave the way for next-generation nanoscale, low-power insulator spintronics devices.
%
%Here, we demonstrate an experimental pathway to detect an interfacial exchange field between insulating ferromagnet EuS and non-magnetic Cu using an all-electrical method, and show that the spin-dependent scattering at the common interface can lead to a significant current-in-plane  magnetoresistance in {Py$|$Cu$|$EuS} trilayer Hall-bar devices.
Here, we demonstrate an experimental pathway to detect an interfacial exchange field between insulating ferromagnet EuS and non-magnetic Cu \mbox{using} magnetoresistance measurements, and show that the spin-dependent scattering at the common interface can lead to a significant current-in-plane magnetoresistance in {Py$|$Cu$|$EuS} trilayer Hall-bar device.
Our experiment suggests that simple magnetoresistance measurements can be used to experimentally detect the interfacial exchange field and thereby the magnetic state of a ferromagnetic insulator.

\end{abstract}
	
	
\maketitle

%Introduction-Importance of interface effects in spintronics
\section{Introduction}

Recently, there has been a strong drive to integrate ferromagnetic insulators into memory and logic devices, requiring all-electrical methods to detect their magnetic state. 
%To fully harness their potential, we need all-electrical methods to detect their magnetic state. One particularly promising approach is to exploit the unique properties that emerge at the interfaces between these materials and other materials.
%
To further miniaturize nanoscale spintronic devices and integrate them with CMOS technology, exploiting interfacial phenomena beyond bulk properties is crucial \cite{2018Manipatruni, 2020Dieny, 2020Hirohata}. 

% mediated by spin-orbit coupling
One of the most studied interface effects is the Rashba-Edelstein effect. The effect originates from the spin-momentum locking and momentum-dependent spin splitting at interfaces with broken inversion symmetry and large spin-orbit coupling. The Rashba-Edelstein effect has been extensively investigated in topological insulators and two-dimensional electron gases \cite{1984Bychkov, Edelstein95, 2013Sanchez, Amin-JAP-Rev, 2016Soumyanarayanan}. 
The electrical characterization of the Rashba-Edelstein effect at interfaces between heavy metals and ferromagnetic insulators is typically achieved by measuring magnetoresistance effects, specifically the spin-Hall magnetoresistance (SMR) \cite{2013-Nakayama} and the unidirectional spin-Hall magnetoresistance (USMR) \cite{Avci2015}. 
Both SMR and USMR rely on the presence of a large spin-orbit coupling effect at the interface between heavy metals and ferromagnetic insulators. Therefore, these effects require expensive heavy metals with high resistance, which are undesirable for energy-efficient device operation \cite{Nguyen2024,Ramesh-rev}.

% mediated by magnetic exchange

Another promising interfacial phenomenon is the interfacial exchange field (IEF) that arises from short-ranged direct spin-spin exchange interactions between adjacent layers of different -- not necessarily magnetic -- materials and thus is highly localized to the common interface, with a decay depth of a couple of nanometers only \cite{Moodera_2007}.
The contemporary technological importance of the interfacial exchange field is mostly based on static equilibrium phenomena. This includes exchange bias in ferromagnet-antiferromagnet systems \cite{1956Meiklejohn}, proximity-induced magnetization \cite{2012Huang_b, 2014Guo_}, and substantial field-free spin splitting (up to several hundred Tesla) in superconductors \cite{1970Meservey, 1991Hao}.
%
%The primary benefit of the interfacial exchange field lies in its ability to be controlled by external magnetic fields, a distinct advantage over the Rashba-Edelstein effect.
The primary benefit of the interfacial exchange field is that the direction of exchange field can be controlled by external magnetic fields without requiring a change in current direction, a distinct advantage over the Rashba-Edelstein effect.
If an all-electrical detection scheme is developed, the interfacial exchange field at the interface between ferromagnetic insulators and nonmagnetic metals can unlock new functionalities in spintronic devices. %, akin to the Rashba-Edelstein effect. 

In this work, we experimentally demonstrate all-electrical detection of interfacial exchange-field using a giant magnetoresistance (GMR)-like device, based on a (conducting ferromagnet)|(non-magnetic~conductor)|(insulating ferromagnet) trilayer in a current-in-plane (CIP) geometry. 
In our case, the conducting ferromagnet is widely used permalloy (Py, Fe$_{0.2}$Ni$_{0.8}$), and the insulating ferromagnet is europium sulfide (EuS).
As intermediate non-magnetic conducting layer we chose highly-conducting copper (Cu), a metal without sizable spin-orbit coupling, which avoids additional spurious spin-Hall or Rashba-Edelstein effects at the Cu$|$Py interface. Although the Rashba-Edelstein effect at the Cu|EuS interfaces can induce magnetoresistance its contribution is expected to be largely temperature-independent, in contrast to the interfacial exchange field (IEF), which is manifested below the Curie temperature of EuS only.

We demonstrate that the interfacial exchange field can be detected via comparative magneto-resistance measurements on trilayer Py$|$Cu$|$EuS Hall-bar devices above and below Curie temperature of EuS. Therefore, simple GMR measurement can provide an all-electrical method to detect IEF.

\section{All electrical detection of the Interfacial Exchange Field}

\begin{figure}[t]
	\centering
	\includegraphics{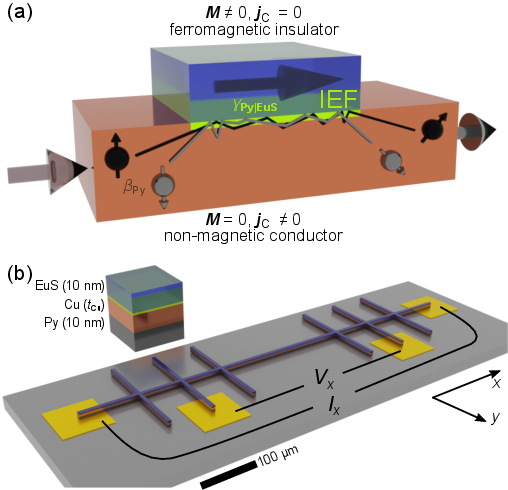}
	\caption{
		\textbf{Magnetoresistance effect due to spin-dependent scattering.}
		(a)~Schematic illustration of spin-dependent reflection of unpolarized electrons from an interfacial exchange field (IEF, green) located at the highly-insulating EuS layer (blue): No charge current flows in EuS, and an unpolarised in-plane charge current $\vec{j}_c$ in the conducting Cu layer (brown) leads to the scattering of the charge carriers at the IEF localized at the Cu$|$EuS interface. 
		If the charge carrier spin (black and silver) is not aligned with the EuS magnetization (large black arrow), the IEF induces spin-dependent scattering, resulting in a magnetoresistance effect akin to current-in-plane giant magnetoresistance observed in all-metallic multilayers.
        (b)~Schematic of the multi-layer Hall-bar device with arm width of \SI{5}{\micro\meter} used to measure the effect of interfacial exchange field on the magnetotransport.
	}
	\label{fig:working-principle}
\end{figure}

One simple method to detect an interfacial exchange field is via a device resembling those used to measure CIP-GMR in metallic trilayers. The working principle of this method is illustrated in Fig. 1(a). 
The magnetoresistance (MR) in these devices can be analyzed using the well-known "two-channel model" \cite{SF-Lee-PRB, Camblong-PRB, Valet-Fert-PRB}. According to this model, the magnetoresistance in all-metal multilayers is determined by the bulk and interface spin-scattering asymmetry coefficients, $\beta$, and $\gamma$, respectively. 
In particular, due to the multiple scattering events at the interface, the current-in-plane-GMR is dominated by spin-dependent scattering at the interfaces\cite{Parkin-APL-PyCu}. 
The interface spin-scattering asymmetry coefficient $\gamma_\text{FM|NM}$ between a ferromagnet (FM) and a normal metal (NM) is defined as\cite{Valet-Fert-PRB, Bass_2007-Rev,BASS2016-Rev}
\begin{equation}
\gamma_\text{FM|NM}=\frac{AR_\text{FM|NM}^{\uparrow}-AR_\text{FM|NM}^{\downarrow }}{AR_\text{FM|NM}^{ \uparrow}+AR_\text{FM|NM}^{\downarrow }}.
\end{equation}

Here $AR_\text{FM|NM}^{\uparrow}$ and $AR_\text{FM|NM}^{\downarrow}$ represent interfacial resistance for the spin-up and spin-down electrons, respectively. The spin-scattering asymmetry coefficient $\gamma_\text{FM|NM}$ thus acts as a source of spin polarization, and in an FM|NM|FM trilayer structure a nonzero $\gamma_\text{FM|NM}$ at both FM|NM interfaces can lead to the observation of a finite magnetoresistance.
%
%The interface-specific resistance is defined by the product of its area and resistance, RA,

%\magenta{[EuS MATERIAL]}
A suitable material to quantify the magnetoresistance originating from interfacial spin-scattering asymmetry at the interface to a highly insulating ferromagnetic insulator (FI), is europium sulfide (EuS). Bulk EuS has a Curie temperature of $T_C^\text{EuS}=\SI{16}{K}$ \cite{1986Mauger}, which allows to activate and deactivate the interface exchange field experimentally by varying the temperature.
EuS is a popular material for superconducting tunnel junctions \cite{1988Moodera,2009Miao, Moodera_2007}, and the strength of the interfacial exchange field from spin-splitting in superconductors in proximity with EuS has been estimated to be about \SI{14}{T} \cite{2016Wei}. 
%
%The influence  (is this equilibrium effects vs. non-equilibrium effects?)
Steady-state effects involving the IEF at the interface to EuS has also been demonstrated as exchange bias \cite{2006Mueller, 2008Volobuev}, the influence of magnetic proximity in various 2D materials \cite{2016Wei, 2017Zhao}, as well as topological insulators \cite{2016Katmis}. 
Recent findings in EuS$|$Pt bilayers furthermore demonstrated the creation of an interfacial non-equilibrium spin accumulation via the spin-Hall magnetoresistance \cite{2020GomezPerez}.

\section{Methods}

%\subsection{Fabrication of Hall cross devices}
Hall-bar devices for magneto-transport measurements were fabricated using mask-less photo-lithography.
To ensure good electrical contact to the bottom Py electrode, and as EuS is insulating and therefore does not allow for post-growth wire-bonded contacts, a two-step lithography procedure was used:
In the first step, gold contact pads were fabricated. For this, photo-resist was spin-coated onto the Si/SiO$_2$ substrates, followed by photo-lithography and development. Then \SI{2}{nm} Ti to promote adhesion and \SI{60}{nm} Au were deposited using electron-beam evaporation, followed by a lift-off process.
In a second lithography step,  photo-resist was spin-coated onto the prepared chip, and Hall-bar devices as shown in Fig.~\ref{fig:working-principle}(b) with \SI{5}{\micro m} leg width and crosses separated by \SI{200}{\micro m}, \SI{320}{\micro m}, and \SI{440}{\micro m} were patterned.

The Py(\SI{10}{nm})|Cu(10~nm or 20~nm)|EuS(\SI{10}{nm}) trilayer thin films and the Hall-bar devices were co-deposited by electron-beam evaporation in an ultra-high vacuum chamber (Kitano Seiki Co., Ltd.) at a base pressure approximately $\SI{2e-8}{Torr}$. The deposition of the different layers was performed without breaking the vacuum.
The substrate stage was maintained at a temperature of \SI{5}{\celsius} during deposition to enhance the crystalline quality and ensure the soft magnetic properties of the EuS thin films \cite{2009Miao, 2018Muduli}.
All films and Hall-bar devices were capped with \SI{5}{nm} AlO$_x$ to prevent degradation, deposited by RF sputtering at room temperature in a separate UHV chamber.
A final ultrasound-assisted lift-off step yielded the completed Hall bar device.

The crystal structure of the EuS thin films was characterized using x-ray diffraction, indicating polycrystalline growth \cite{PRB-2024-EuS, 2018Muduli}. Surface morphology measurements via atomic force microscopy (AFM) verified smooth films with a root-mean-square (rms) roughness of \SI{0.3}{nm} \cite{PRB-2024-EuS}. 

The EuS thin films are highly insulating, with resistances measured in a two-point configuration much larger than \SI{50}{\mega\ohm}, exceeding the limit of the digital multimeter (Fluke). 
Previous publications estimate the resistance of EuS thin films from four-point measurements to  \SI{68}{\mega\ohm} at \SI{300}{K}, increasing to \SI{10}{T\ohm} at \SI{200}{K} \cite{2020GomezPerez}. As all our measurements have been performed below \SI{50}{K}, we therefore can safely assume that any spurious current through the EuS layer can be neglected (see Supplemental Materials Sec.~S1 about the discussion of current shunting).

Temperature- and field-dependent magnetization profiles of multilayer films were measured using DC magnetometry in a Quantum Design MPMS magnetometer. Both longitudinal and transverse magneto-resistances were measured using the four-probe method in a custom-built flow cryostat, using a phase-sensitive lock-in technique with an applied AC current of $I=$\SI{1}{\micro A}-\SI{10}{\micro A} modulated at a frequency of $f=\SI{173}{Hz}$.
	
\section{Results}
	
\begin{figure}[tb]
		\centering
		\includegraphics{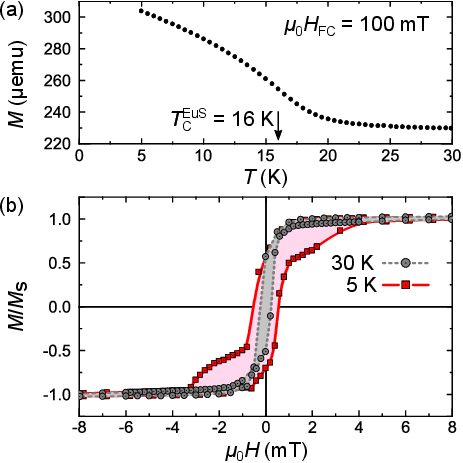}
		\caption{
			\textbf{Magnetometry of Py$|$Cu(\SI{10}{nm})$|$EuS sample.}
			(a)~Temperature-dependent magnetization, measured upon cooling with an in-plane magnetic field $\mu_0H_\text{FC}=\SI{100}{mT}$. Below $T_C^\mathrm{EuS}\approx\SI{16}{K}$ EuS is ferromagnetic.
			(b)~Normalized hysteresis loops $M(H)$ measured at \SI{30}{K} (gray circles) and \SI{5}{K} (red squares) with an in-plane field.  The high-temperature easy-axis loop has a coercive field of $H_c(\SI{30}{K})\approx\SI{0.5}{mT}$. As EuS becomes ferromagnetic at low temperatures, the coercive field increases $H_c(\SI{5}{K})\approx\SI{1}{mT}$ and the loop broadens, indicating interactions between the two magnetic layers and formation of compensated and non-collinear spin configurations at low fields. }
		\label{fig:magnetization}
\end{figure}

\begin{figure}[bt]
	\centering
	\includegraphics{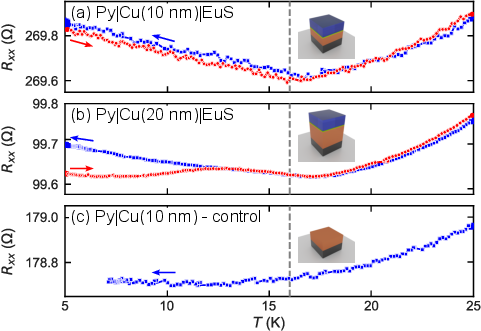}
	\caption{%
		\textbf{Temperature-dependent longitudinal resistance} of multi-layer structures consisting of permalloy (gray), copper (orange), and EuS (blue), as indicated by inset schematics.
		(a,b)~For Py|Cu|EuS devices, $R_{xx}(T)$ shows a clear upturn below $T_C^\text{EuS}$ (dashed vertical line). The $R_{xx}(T)$ exhibits thermal hysteresis during cooling (blue squares) and heating (red circles) cycles at zero field.
		(c)~The control Py|Cu device exhibits a low-temperature resistance saturation without any upturn or hysteresis.
		}
	\label{sfig:magnetotransport_T} % S2
\end{figure}

 \begin{figure}[tb]
	\centering
	\includegraphics{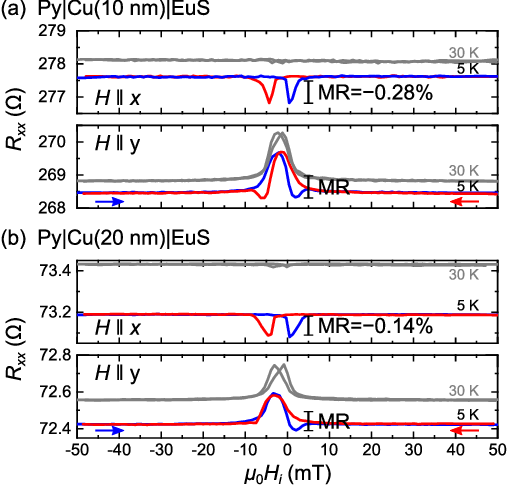}
	\caption{%
		\textbf{Longitudinal magnetoresistance}, measured with different in-plane fields at \SI{5}{K}, where both Py and EuS are ferromagnetic (blue and red lines for field up and down sweep, respectively), and at \SI{30}{K} (only Py ferromagnetic, gray lines).
		(a)~$R_{xx}(H_x)$ and $R_{xx}(H_y)$ for the Py|Cu(\SI{10}{nm})|EuS sample. $R_{xx}(H_x)$ exhibits a drop in resistance at low temperatures due to the GMR-like effect arising from spin scattering between the rotated Py and EuS magnetization at low fields. For $R_{xx}(H_y)$ an additional AMR arises from non-collinear spin textures in Py, and therefore persists in the \SI{30}{K} control experiment.
		(b)~$R_{xx}(H_x)$ and $R_{xx}(H_y)$ measured for the Py|Cu(\SI{20}{nm})|EuS sample. Its behaviour is equivalent to that of the other sample, albeit with a lower resistance due to the thicker Cu layer, and a lower value of $\text{MR}$.
	}
	\label{fig:long_magnetoresistance_H}
\end{figure}

As shown in Fig.~\ref{fig:magnetization}(a), the field-cooled magnetization in the Py$|$Cu$|$EuS trilayer shows an increase below \SI{16}{K}, which coincides with the ferromagnetic ordering temperature $T_C^\text{EuS}$ of high-quality EuS films \cite{Moodera_2007,2018Muduli}. The non-zero baseline magnetization of about \SI{230}{\micro emu} is due to Py, which has a much larger Curie temperature ($T_C^\text{Py}\approx\SI{800}{K}$). 
Like Py, EuS films exhibit an in-plane magnetic anisotropy (Supplemental Materials Sec.~S2). 

Figure~\ref{fig:magnetization}(b) shows normalized hysteresis loops $M(H)/M_\text{sat}$ of the Py|Cu(\SI{10}{nm})|EuS trilayer for in-plane field $H$, measured at \SI{30}{K} and \SI{5}{K} (gray circles and red squares, respectively).
Compared to the soft-magnetic hysteresis loop at \SI{30}{K}, where only Py is ferromagnetic, the low-temperature loop is broader due to the interaction between the ferromagnetic layers, however, a low field of \SI{4}{mT} is sufficient to achieve complete switching and full collinear alignment of Py and EuS magnetisation.
The more complex hysteresis loop at low temperatures, with a slightly pinched behavior and "wings" at higher fields, is reminiscent of the hysteresis of vortex states in soft magnetic films, and indicates spin curling between the Py and EuS magnetization. We furthermore demonstrated a non-collinear alignment at the interface between Py|EuS bilayers using polarised neutron reflectometry with polariation analysis \cite{202xCaruana}.
With the observation of collinear and non-collinear magnetic configurations in Py|Cu|EuS trilayers, we fulfill a key requirement for the observation of a GMR signal related to spin-dependent reflection from the IEF. 
%\section{Temperature-dependent magnetoresistance}

We now discuss the longitudinal magnetoresistance $R_{xx}=V_{x}/{I_x}$ , with in the measurement geometry shown in Fig.~\ref{fig:working-principle} (b). Here, a voltage $V_{x}$ is measured parallel to the charge current $\vec{j}_c\parallel x$ and with in-plane magnetic fields applied either along $x$ or $y$ directions.

Fig.~\ref{sfig:magnetotransport_T} summarizes the temperature dependence of the longitudinal resistance $R_{xx}(T)$, with measurements taken first upon heating from \SI{5}{K} to \SI{50}{K} (red circles), and then cooling back to \SI{5}{K} (blue squares), all in zero magnetic field.
For both Py|Cu($t_\text{Cu}$)|EuS devices the zero-field curves show a clear upturn at $T_C^\text{EuS}\approx\SI{16}{K}$, Fig.~\ref{sfig:magnetotransport_T}(a,b), below which EuS becomes ferromagnetic and spin-dependent scattering from the IEF comes into play.
The longitudinal resistance $R_{xx}(T)$ is also found to exhibit thermal hysteresis below $T_C^\text{EuS}$ during cooling (blue squares) and heating (red circles) cycles at zero field. 
%This is likely related to the emergence of complex multi-domain states. 
The thermal hysteresis observed in both the $t_\text{Cu}$=\SI{10}{nm} and \SI{20}{nm} samples demonstrated a dependence on the magnitude and direction of the applied magnetic field, exhibiting either an amplification or a suppression effect. This implies that thermal hysteresis is intrinsically linked to the complex multi-domain state of the EuS \cite{2014Idzuchi} and Py layers (see Supplemental Material Sec.~3).
%The hysteresis is more pronounced for the sample with the $t_\mathrm{Cu}=\SI{20}{nm}$ thick Cu interlayer [Fig.~\ref{sfig:magnetotransport_T}(b)], probably due to the weakening of the magnetic coupling between Py and EuS layer.
%
%
Furthermore, the clear kink and upturn of the resistance at $T_C^\text{EuS}$ is notably absent in a Py|Cu reference device, as shown in Fig.~\ref{sfig:magnetotransport_T}(c), where the temperature dependence of $R_{xx}(T)$ follows regular metallic behavior, saturating to a constant resistance value below \SI{10}{K}.
Thus, we observe a clear variation of $R_{xx}(T)$ in the Py|Cu|EuS devices at and below $T_C^\text{EuS}$, providing direct evidence for the onset of an interfacial-exchange-field-mediated spin-dependent scattering at the Cu|EuS interface as EuS turns ferromagnetic.

In order to accurately quantify the GMR-like resistance change associated with spin-dependent scattering, while mitigating the influence of magnetoresistive effects originating from the Py layer, we will conduct measurements both above and below the Curie temperature of EuS, $T_C^\text{EuS}$. By employing this approach, we can discriminate between the anisotropic magnetoresistance (AMR) originating from the Py layer only and the GMR-like contributions coming from the Py|Cu|EuS trilayer
as we fully suppress the AMR contribution for the measurements with current and field parallel, i.e., $j\parallel H\parallel x$ (also see Supplemental Materials Sec.~S3).

Figure~\ref{fig:long_magnetoresistance_H} shows the longitudinal magnetoresistance $R_{xx}(H_i)$ ($i=x,y$) for the two Py|Cu($t_\text{Cu}$)|EuS samples with $t_\text{Cu}=\SI{10}{nm}$ and \SI{20}{nm} shown in Fig.~\ref{fig:long_magnetoresistance_H}(a) and (b), respectively.
In particular, the $R_{xx}(H_x)$ measurements at \SI{5}{K} (red and blue lines) reveal distinct dips centered around \SI{\pm3}{mT} for both samples. These dips are not observed in the higher-temperature \SI{30}{K} data (gray lines), providing clear evidence of GMR-like resistance changes occurring below $T_C^\text{EuS}$.
We estimated the MR using the formula
\begin{equation}
		\text{MR} = \frac{R_{\uparrow\downarrow} -R_{\uparrow\uparrow}}{\min\left(R_{\uparrow\downarrow},\, R_{\uparrow\uparrow}\right)} \, ,
		\label{eq:MR}
\end{equation}
where $R_{\uparrow\uparrow}$ and $R_{\uparrow\downarrow}$ denote the resistance in the state with parallel and antiparallel alignment of the two ferromagnets, respectively.
From the low-temperature $R_{xx}(H_x)$ curves shown in Fig.~\ref{fig:long_magnetoresistance_H} the magnetoresistance $\text{MR}$ is determined to be $\text{MR}_{\SI{10}{nm}}=\SI{-0.28}{\percent}$ for the $t_\text{Cu}=\SI{10}{nm}$ device ($\text{MR}_{\SI{20}{nm}}=\SI{-0.14}{\percent}$). 
Comparable GMR values are obtained for $R_{xx}(H_y)$, however, the uncertainty in these measurements is larger due to the presence of additional AMR contributions originating from the complex spin configuration in the Py layer (see Supplemental Materials Sec.~S3). This AMR effect is particularly pronounced near zero field and can still be observed above $T_C^\mathrm{EuS}$ at \SI{30}{K}.

Beyond the IEF, a strong Rashba spin-orbit coupling is also present at the EuS|Cu interface, as evidenced by our recent room-temperature spin-torque ferromagnetic resonance (STFMR) experiments \cite{PRB-2024-EuS}. While the presence of such a strong Rashba effect might suggest a potential contribution to the observed GMR-like resistance, it is important to note that this effect persists even in the paramagnetic state of EuS above $T_C^\text{EuS}$ up to room temperature. 
Furthermore, recent studies on the Rashba-Edelstein effect on EuS|Pt bilayers reported maximum spin Hall magentoresistance amplitudes of \num{3e-5} \cite{2020GomezPerez}, which is about two orders of magnitude smaller than our observed MR values of $\approx0.3\%$.
Consequently, the GMR-like effect, which here is exclusively observed below $T_C^\text{EuS}$ in Py|Cu|EuS thin films, cannot be attributed to the Rashba-Edelstein effect and therefore must be related to the IEF only.

\section{Discussion}

The observation of the finite GMR-like signals suggests a nonzero spin-scattering asymmetry.
In our analysis, we have disregarded the influence of bulk spin-asymmetry coefficients, $\beta_\textrm{Py}$ and $\beta_\textrm{Cu}$. This is based on previous experimental studies on Py|Cu|Py CIP-GMR devices, which suggest that the interfacial spin-dependent anisotropy parameter $\gamma_\text{FM|NM}$ plays a more significant role than bulk spin-asymmetry coefficients in these types of CIP-GMR trilayer structures, due to the many scattering events at the interface \cite{Parkin-APL-PyCu}.

This leaves the parameter $\gamma_\text{FM|NM}$ at both the Py|Cu and EuS|Cu interfaces as source for the spin-scattering asymmetry. For the Py|Cu interface, this value is known from previous works to be $\gamma_\text{Py|Cu} = 0.81$ at \SI{4.2}{K} \cite{Yang-Py-B,BASS2016-Rev}. 
However, the spin-dependent anisotropy parameters $\gamma_\text{EuS|Cu}$ for the EuS|Cu interface have not been investigated neither experimentally nor theoretically. We observed a negative MR, which suggests $\gamma_\text{EuS|Cu}$ is nonzero and negative. The $\gamma_\text{FM|NM}$, which is related to the electronic structure at each interface, has shown to be negative for many other FM|NM interfaces as well \cite{1999Vouille, 2004Dijken, 2006Yang}. 
In our experiment, it was not feasible to quantitatively estimate the value of $\gamma_\text{EuS|Cu}$. To obtain a precise quantitative estimate of $\gamma_\text{EuS|Cu}$, further experiments involving a systematic variation of both the Cu and Py layer thicknesses are required \cite{BASS2016-Rev, Yang-Py-B,PhysRevB.58.12230}.

% I slightly shortened this section
Our experiments suggest that by performing simple MR measurements on FI|NM|FM trilayers, we can develop an all-electrical technique to detect the IEF at the FI|NM interface, which opens new avenues for insulatronic devices. Furthermore, by optimizing the interface spin-asymmetry coefficients $\gamma_\text{FI|NM}$, we can enhance the MR signal in these devices. 
Future experiments should consider a more systematic series of devices varying Cu spacer thicknesses to disentangle how the spin diffusion length in Cu or interfacial spin-memory loss affects the MR magnitude.
In our case, we are sensitive to the effect in-plane magnetisation. Recently, a GMR-like effect arising from spin-dependent electron scattering at the FI|NM interface was observed for a terbium iron garnet (TbIG)-based  TbIG|Cu|TbCo trilayer to detect a out-of-plane magnetisation reversal in TbIG \cite{Damerio2024}. %It was shown that GMR-like measurements can be employed to detect perpendicular magnetization reversal of TbIG. 

%Our proposed method opens up new avenues for insulatronics devices. 
%
Beyond ferromagnetic insulators, we anticipate that this approach can be extended to detect the interfacial exchange field at the interface between antiferromagnetic insulators and metals, providing a straightforward all-electrical method to read out the magnetic state of antiferromagnets.

\section{Conclusions}
In this work, we present a novel approach to detect the interfacial exchange field using a CIP-GMR device comprising one metallic and one insulating ferromagnet. As a proof-of-concept, we fabricated GMR-like Py|Cu|EuS trilayer thin films and Hall-bar devices.
Magnetization measurements confirmed the parallel and antiparallel alignment of the Py and EuS layers below the Curie temperature of EuS. Magnetoresistance measurements on the Py|Cu|EuS Hall-bar devices revealed a CIP-GMR-like resistance, dependent on the relative magnetization direction of the Py and EuS layers. Given its highly-insulating nature, no current flows through the EuS layer, and we attribute this GMR-like signal to spin-dependent scattering induced by the IEF at the EuS|Cu interface. Although we achieved only a modest MR of -0.28$\%$ in Py|Cu(10~nm)|EuS trilayers, optimization via interface engineering can significantly enhance this value.

Our results therefore show that simple MR measurements offer a straightforward, all-electrical method to detect the IEF at the surface of ferromagnetic insulators, opening up exciting opportunities for insulating spintronics devices that operate at low current densities and exhibit additional magnetic reconfigurability and non-volatility. Future research should focus on estimating the interface spin-dependent anisotropy parameter $\gamma_\text{IF|NM}$ at interfaces to various highly insulating ferro-, ferri- and antiferromagnets, including garnets \cite{2018GomezPerez, 2019Velez, Damerio2024}, which also would allow for room-temperature operation.

\section*{Data Availability}
The data set supporting the data presented in this work can be found in the zenodo archive with DOI \href{https://doi.org/10.5281/zenodo.14840166}{10.5281/zenodo.14840166}.

\begin{acknowledgments}
		
We thank Touru Yamauchi for giving access to the MPMS facility at ISSP, and Jonathan Leliaert for helpful suggestions regarding the implementation of the micromagnetic simulations.

This work was supported by a Grant-in-Aid for Scientific Research on Innovative Area, Nano Spin Conversion Science, MEXT, Japan (Grant No.~26103002).
N.L.~received funding from the Japan Society for the Promotion of Science as JSPS International Research Fellow (grant ID PE18701), the European Research Council (ERC) under the European Union’s Horizon 2020 research and innovation program [Marie Sk{\l}odowska Curie Grant Agreement No. 844304 (LICONAMCO)], as well as by a UKRI Future Leader Fellowship [grant number MR/X033910/1 (LIONESS)].
% FY2018 JSPS Postdoctoral Fellowship for Reseach in Japan (Short-term)
% ID No. PE18701
%"JSPS International Research Fellows"
%
P.K.M.~acknowledges funding from IIT Madras (Grant Numbers IP21221798PHNFSC008989 and RF21221392PHNFIG008989) and Science and Engineering Research Board (SERB), India Grant No.~SRG/2022/000438.
M.X. would like to thank support from JSPS through the “Research Program for Young Scientists” (no. 19 J21720) and the RIKEN IPA Program.

\end{acknowledgments}

\clearpage \newpage
%\newrefsection

\beginsupplement
%\newrefsection % separate bibliography for main text and supplement
\title{\papertitle\linebreak Supplemental Material}
\maketitle

\section{Current shunting in trilayers}
\label{ssec:shunting} % S1

Typical resistances of EuS films at low temperatures are very large (see Methods), and we can safely disregard any charge conduction through the EuS layer.
In the Py|Cu|EuS trilayer Hall bars therefore the net charge current $I_T$ flows parallel through both the Cu and Py layer, so that $I_T=I_\mathrm{Py} + I_\mathrm{Cu}$ (current shunting). 
Due to the partial shunting in the magnetic Py layer, the charge current $I_\text{Cu}$ in the Cu layer is also spin-polarised with a spatially varying polarization direction determined by the local moment at the Py$|$Cu interface.

The ratio $\eta$ between currents $I_\mathrm{Py}$ and $I_\mathrm{Cu}$ is inversely proportional to the resistances of the respective layers, 
\begin{equation}
	\eta 
	= \frac{ I_\mathrm{Cu} }{ I_\mathrm{Py} } 
	= \frac{ R_\mathrm{Py} }{ R_\mathrm{Cu} } 
	= \frac{ \rho_\mathrm{Py}}{ \rho_\mathrm{Cu}} \, \frac{ t_\mathrm{Cu} }{ t_\mathrm{Py} }  
	\, ,
	\label{eq:current_ratio}
\end{equation}
and the fraction of the respective current flowing through both the Cu and Py layer then can be written as 
\begin{equation}
	\frac{I_\mathrm{Cu}}{I_T} = \frac{\eta}{1 + \eta} 
	\qquad \mathrm{and} \qquad 
	\frac{I_\mathrm{Py}}{I_T} = \frac{1}{1 + \eta} 	
	\, .
	\label{eq:current_fraction}
\end{equation}

The resistances in the wires is given by $R_i=\rho_i \frac{l_i}{w_i t_i}$
%\begin{equation}
%    R_i=\rho_i \frac{l_i}{w_i t_i} \, ,
%\end{equation}
with material-dependent resistivity $\rho_i$ ($i$ = Py, Cu) and $l_i$, $w_i$, and $t_i$ as lengths, widths, and thicknesses of the respective layer of the wire. 
As the length and width of the wires are equal, ${l_\mathrm{Cu}=l_\mathrm{Py}}$ and ${w_\mathrm{Cu}=w_\mathrm{Py}}$, only the layer thicknesses appear in Eq.~(\ref{eq:current_ratio}).

At \SI{10}{K} the resistivities are $\rho_\mathrm{Cu}\approx\SI{2}{\micro\Omega\,cm}$ and $\rho_\mathrm{Py}\approx\SI{32}{\micro\Omega\,cm}$ \cite{2018Muduli}.
With these values we obtain current ratios ${\eta(t_\mathrm{Cu}=\SI{10}{nm})\approx 16}$ and ${\eta(t_\mathrm{Cu}=\SI{20}{nm})\approx 32}$ for the Py$|$Cu$|$EuS devices used in this work.
We thus estimate that \SI{94}{\percent} of the net charge current flows through the trilayer device with Cu layer thickness $t_\mathrm{Cu}=\SI{10}{nm}$ (respectively \SI{97}{\percent} for $t_\mathrm{Cu}=\SI{20}{nm}$).

\begin{figure}[b]
	\centering
	\includegraphics[width=67mm]{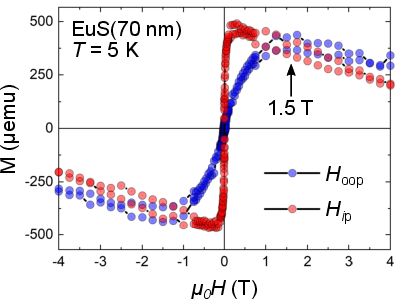}
	\caption{%
		\textbf{Magnetic anisotropy.}
		Hysteresis loops of a \SI{70}{nm}-thick EuS film measured at \SI{5}{K} with in-plane (closed red circles) and out-of-plane (closed blue circles) fields, showing easy-plane anisotropy. The negative slope at high fields originates from the diamagnetic contribution of the Si/SiO$_2$ substrate.
	} 
	\label{sfig:magnetic_anisotropy} % S1
\end{figure}

\section{Magnetic anisotropy of EuS}
\label{ssec:magnetic_anisotropy}

Below the Curie temperature of $T_C^\mathrm{EuS}\approx\SI{16}{K}$, EuS is a soft ferromagnet with an in-plane magnetic anisotropy, as demonstrated by hysteresis loops shown in Fig.~S1. Here,  a small field of \SI{4}{mT} is sufficient to switch the in-plane magnetization, whereas a much higher field of about \SI{1.5}{T} is necessary to saturate the magnetization out-of-plane.

\section{Micromagnetic Simulations}
\label{ssec:mumax}

\begin{figure}[tb]
	\includegraphics{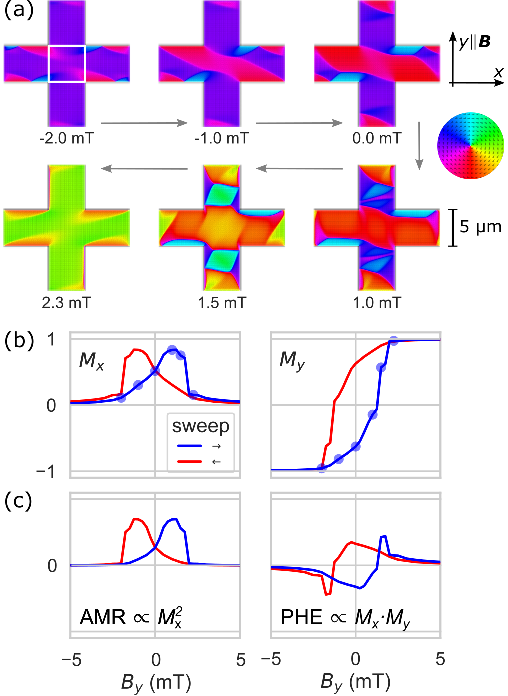}
	\caption{%
		\textbf{Micromagnetic behaviour of Py Hall cross,} from mumax3 simulations. 
		(a)~Micromagnetic configurations at low fields.
		(b)~Average magnetization $M_x$ and $M_y$ in the central region of the Hall cross [white outline in first image of (a)], blue circles correspond to images shown in (a).
		(c)~Expected shape of possible magnetoresistance contributions due to field-dependent non-collinear spin textures (for perpendicular $\vec{j}_c\parallel x$ and $\vec{B}\parallel y$).
	} \label{sfig:mumax} %S3
\end{figure}

%\subsection{Micromagnetic simulations}
Finite-difference micromagnetic simulations were implemented using the GPU-accelerated software package MuMax3 \cite{2014Vansteenkiste}. The geometry of a permalloy Hall cross of \SI{5}{\micro m} leg width, \SI{17.5}{\micro m} leg length, and a thickness of $t=\SI{10}{nm}$ was discretised with cuboid cells of \SI{5}{nm} x \SI{5}{nm} x \SI{10}{nm} size. Material constants of bulk permalloy were used, with saturation magnetization ${M_\text{sat}=\SI{860}{kA/m}}$, exchange constant ${A_\text{ex}=\SI{13}{pJ/m}}$, and vanishing magneto-crystalline anisotropy $K=0$. The equilibrium magnetization configuration was simulated for static magnetic fields applied along the vertical $y$ direction, with a small field of \SI{0.1}{mT} applied along the horizontal $x$ direction to break the symmetry. To emulate the response of an infinite wire in either direction, the magnetic surface charges at the bounding box boundary were compensated.

Figure~S3(a) shows snapshots of the magnetisation at low fields, and (b) plots magnetization components in the central region of the Hall cross [see white outline in (a)] under application vertical fields $B_y$: While $M_y$ shows a conventional hysteresis loop, due to the soft magnetic behaviour sizable in-plane contributions $M_x$ emerge.

The complex non-uniform spin textures within the Py Hall cross allow for additional magnetoresistance effects to emerge, proportional to specific products of the magnetization components, as shown in Fig.~S3(c). These contributions make quantification of the GMR due to the interfacial exchange field more difficult.

In the case of the longitudinal magneto\-resistance $R_{xx}$, the response is dominated by the anisotropic magneto\-resistance (AMR) \cite{1975McGuire} with
\begin{equation}
	R_\text{AMR}  = R_\bot + (R_\parallel - R_\bot) M^2_x \, ,
	\label{seq:AMR}
\end{equation}
where $R_\bot$ and $R_\parallel$ are resistance in the configuration $\vec{j}\bot\vec{M}$ and $\vec{j}\parallel\vec{M} $, respectively.
In the results shown in Fig.~4 of the main manuscript, the AMR leads to the peaks when the current is perpendicular to the field $(\vec{j}_c\parallel x)\perp H_y$, and the field approaches zero $H_y\rightarrow 0$, which are especially noticeable at \SI{30}{K}, where EuS is paramagnetic.

The transverse magnetoresistance $R_{xy}$ is \cite{1975McGuire, 1954Goldberg}
\begin{equation}
	R_{xy}
	= \frac{\mu_0 R_0 }{t} H_z  + \frac{\mu _0 R_S }{t} M_z  + \frac{\Delta R}{t} M_x M_y  \, .
	\label{seq:TMR}
\end{equation}
Here, $R_0$ and $R_S$ are the coefficients for the ordinary Hall effect (OHE) and anomalous Hall effect (AHE), respectively. The third term corresponds to the planar Hall effect (PHE), which is related to the AMR of Eq.~(\ref{seq:AMR}) via the coefficient ${\Delta R=(R_\parallel- R_\bot)}$.
As shown in Fig.~S3(b), due to the dependence of the PHE on the product $M_xM_y$ we expect an asymmetric  $R_{xy}(\mu_0H_{x,y})$ profile upon sweeping the in-plane magnetic field.

\end{document}